\documentclass{llncs}

\begin{document}

\title{Data Language Specification via Terminal Attribution}

\author{Alexander Sakharov, Timothy Sakharov}

\institute{}

\maketitle

\begin{abstract}
Unstructured data have to be parsed in order to become usable. The complexity of grammar notations and the difficulty of grammar debugging limit the use of parsers for data preprocessing. We introduce a notation in which grammars are defined by simply dividing terminals into predefined classes and then splitting elements of some classes into multiple layered sub-groups. These LL(1) grammars are designed for data languages. They simplify the task of developing data parsers.
\end{abstract}

\section{Introduction}
Most data are unstructured. Parsing unstructured data opens up opportunities for further processing, querying, and extracting knowledge from the data, as well as loading data into databases or publishing web pages. Many software professionals are involved in the development of data parsers without even realizing that the code they wrote is actually a hard-coded parser. Hard-coded parsing is a typical step in big data preprocessing. These hard-coded data parsers require software updates with every change in data format. The emergence of standardized data notations such as XML and JSON is essentially a response to the technical difficulties associated with parsing data, but the reality is that only a small percentage of data conform to these standards.

Context-free grammars (CFG) \cite{Aho06} are an excellent mechanism for specifying the syntax of programming languages, but they are rarely used in data preprocessing. Few software developers are familiar with CFGs. In most cases, it is still easier to hard-code an ad hoc data parser than to write and debug a CFG grammar. Understanding how to create a grammar for predictive top-down parsing or a grammar without conflicts for bottom-up parsing requires some deep understanding of the theory of parsing. Basically, CFGs are not for everyone but only for gurus in the domain of formal languages and compilers. Over the course of time, several alternative grammar notations were developed. With the exception of regular expressions, none of these alternatives really simplified the task of creating and debugging grammars. 

The output of parsing both programming and data languages is the same: it is an abstract syntax tree (AST) \cite{Aho06} which contains syntactic information extracted from the source. Nonetheless, there are some fundamental differences between parsing programming languages and parsing data. Data languages mostly consist of aggregation constructs and references. The former represent structures with named fields or sets including maps. i.e. key-value pairs. A grammar defining a programming language should be very constraining. It should disallow inappropriate strings of symbols because language semantics should be applicable to any well-formed program according to the grammar defining the language. 

The situation is drastically different with the parsing of data. The goal of parsing data is not to verify the conformance, but to build a rich parse tree, subsequently extract and tag pieces of information, and possibly derive relationships among these pieces based on the structure of the parse tree. Almost always, some portions of data have an incorrect format or somehow diverge from any given standard. Therefore, grammars for defining the syntax of data should be inclusive in order to avoid undesirable exceptions when processing these data. In contrast to programming languages, data formats are plentiful and evolve all the time. It is important, especially for big data, to be able to easily modify data grammars without the danger of compromising their properties. It is also important to be able to parse data using an incomplete grammar because the exact syntax of big data may not be known in advance.

An adequate notation for defining data languages should be much simpler than CFGs. It should be on par with regular expressions in terms of comprehensibility. Unfortunately, regular expressions themselves are not a good choice for defining data languages because of their limited expressiveness and because they do not help build informative parse trees. The use of such notation should not require sophisticated tools for parser generation, and parsing should be feasible in linear time. A notation that satisfies the above criteria was introduced in \cite{Sakharov15}. It is called tier grammars. This paper provides a more detailed description of tier grammar properties. 

Tier grammars have no nonterminals, no grammar productions, and no formulas. A language is defined by simply dividing terminals into predefined classes. Each class has its role. Some classes are split into multiple tiers, i.e. layered sub-groups. Note that the choice of terminal classes in this notation is not motivated by theoretical considerations but rather is driven by the intent to cover more constructs used in practice while having a clear meaning of every terminal class. Tier grammars define a subset of LL(1) languages, which makes predictive parsing possible \cite{Aho06}. Tier languages are unambiguous. They are devised to be very inclusive. We give a simple characterization of strings belonging to these languages. This notation is rich enough for specifying data formats of various kinds of documents including machine-generated documents such as log files. Our notation facilitates the definition of constructs representing data aggregates and references.  

\section{Definition of Tier Languages}

Following the tradition for programming languages, it is assumed that lexical analysis using regular expressions is done before parsing. The output of lexical analysis is a sequence of tokens whose names are terminals for parsing. As usual, the longest lexeme is selected in case of conflicts \cite{Aho06}. If the syntax is known for portions of the input, then regular expressions are also used to select these fragments before parsing them. 

Suppose the set of terminals T is a union of disjoint sets $T_1$, $T_2$, $T_3$, $T_4$, $T_5$, $T_6$, $T_7$. $T_1$ is the set of base terminals. Terminals from $T_2$ and $T_3$ define bracketed constructs. Terminals from $T_2$ are opening brackets, and terminals from $T_3$ are closing ones. Terminals from $T_4$ are called markers. These terminals are split into groups by their priority. Their role is to serve as delimiters that combine items left and right to them in groups. 

Terminals from $T_5$ are called postfixes. They are postfix operators. Terminals from $T_6$ are called prefixes. They are unary prefix operators. Terminals from $T_7$ are connectives that serve as binary operators in expressions or separators like in the comma-separated values format. Prefixes, postfixes, and connectives are also split into disjoint groups by their priority. They share the range of priorities but only one kind of terminal is allowed for a given priority. Let $q$ be the highest priority for markers and $k$ be the highest priority for postfixes, prefixes, and connectives. We use $\underline i$ to denote the number of distinct markers, postfixes, prefixes, or connectives of priority $i$.

Tier language $\Lambda(T)$ for family of terminal classes $T = \{ T_1$, $T_2$, $T_3$, $T_4$, $T_5$, $T_6$, $T_7 \}$ is defined recursively by the following rules. Understanding these rules does not require any knowledge of CFGs, but tier languages can still be expressed via CFGs. We give CFG productions along with the rules in order to demonstrate how the rules map to them. $S$ will denote the start nonterminal of the corresponding CFG. Symbol $\epsilon$ will denote the empty string. $T_{4i}, T_{5i}, T_{6i}, T_{7i}$ will denote respective terminals of priority $i$. Note that only one of $T_{5i}, T_{6i}, T_{7i}$ may be non-empty for any $i$.

1. If $b \in T_1 = \{ b_1,...,b_{\underline b} \}$, then $b \in \Lambda(T)$.

\noindent
$A \rightarrow b_1 | ... | b_{\underline b}$

2. If $a \in \Lambda(T),  r \in T_2 = \{ r_1,...,r_{\underline r} \},  e \in T_3 = \{ e_1,...,e_{\underline e} \}$, then  $r a e \in  \Lambda(T)$.

\noindent
$B \rightarrow F S H$

\noindent
$F \rightarrow r_1 | ... | r_{\underline r}$

\noindent
$H \rightarrow e_1 | ... | e_{\underline e}$

3. Let either $c_1,...,c_n \in T_{7i} = \{ c_{i1},...,c_{i \underline i} \}$ (connective), $p \in T_{6i} = \{ p_{i1},...,p_{i \underline i} \}$ (prefix), or $s \in T_{5i} = \{ s_{i1},...,s_{i \underline i} \}$ (postfix). If $a_1,...,a_n, a_{n+1} \in \Lambda(T)$, $a_1,...a_n$, $a_{n+1}$ are defined by rules 1, 2, or this rule for terminals of higher priority, $a_0 \in \Lambda(T), a_0$ is defined by rules 1, 2, or this rule for terminals of the same or higher priority, then $a_1 c_1 a_2 c_2 ... a_n c_n a_{n+1} \in \Lambda(T)$, $p a_0 \in \Lambda(T)$, or $a_1 s \in \Lambda(T)$.

\noindent
$C \rightarrow A | B$

postfix:

\noindent
$E_i \rightarrow E_{i+1} G_i$ (for $i = 1,...,k-1$)

\noindent 
$E_k \rightarrow C G_k$

\noindent
$G_i \rightarrow \epsilon | s_{i1} | ... | s_{i \underline i}$

prefix:

\noindent
$E_i \rightarrow E_{i+1} | p_{i1} E_i | ... | p_{i \underline i} E_i$ (for $i = 1,...,k-1$)

\noindent
$E_k \rightarrow C | p_{k1} E_k | ... | p_{k \underline k} E_k$

connective:

\noindent
$E_i \rightarrow E_{i+1} L_i$ (for $i = 1,...,k-1$)

\noindent
$E_k \rightarrow C L_k$

\noindent
$L_i \rightarrow \epsilon | c_{i1} E_{i+1} L_i | ... | c_{i \underline i} E_{i+1} L_i$ (for $i = 1,...,k-1$)

\noindent
$L_k \rightarrow \epsilon | c_{k1} C L_k | ... | c_{k \underline k} C L_k$ 

4. If $m_1,...,m_n \in T_{4i} = \{ m_{i1},...,m_{i \underline i} \}$, $a_0,a_1,...,a_n \in \Lambda(T)$, then $\epsilon \in \Lambda(T)$, $a_1 ... a_n \in \Lambda(T)$, $a_0 m_1 a_1 ... m_n a_n \in \Lambda(T)$ provided that this string follows the beginning of the input string, or a terminal from $T_2$ or a marker of lower priority, and precedes the end of the input string, or a terminal from $T_3$ or a marker of lower priority.

\noindent
$Q_i \rightarrow Q_{i+1} R_i$ (for $i = 1,...,q-1$)

\noindent
$Q_q \rightarrow D R_q$

\noindent
$R_i \rightarrow \epsilon | m_{i1} Q_{i+1} R_i | ... | m_{i \underline i} Q_{i+1} R_i$ (for $i = 1,...,q-1$)

\noindent
$R_q \rightarrow \epsilon | m_{q1} D R_q | ... | m_{q \underline q} D R_q$

\noindent
$D \rightarrow \epsilon | E_1 D$

Now we only need to add one more production to complete the definition of the corresponding CFG: 
$S \rightarrow Q_1$.
The above context-free productions have to be slightly modified when some terminal sets are empty. In case the sets of connectives, postfixes, and prefixes are all empty:
$E_1 \rightarrow C$.
In case the set of markers is empty:
$Q_1 \rightarrow D$.

These terminal classes  are suitable for various representations of data aggregates and references: prefixes, postfixes, and connectives for named fields in structures; brackets and markers for structures, sets, and maps; connectives for key/value pairs and for separating set elements; prefixes and brackets for references. Rule applications define parse trees for tier languages. Applications of Rule 1 constitute the terminal nodes of these parse trees. Every application of all other rules corresponds to a nonterminal node of the parse tree.

\section{Predictive Parsing}

Let $\alpha$ and $\beta$ denote strings of terminals and nonterminals. A CFG is LL(1) if and only if the following holds for every two productions $A \rightarrow \alpha$, $A \rightarrow \beta$ \cite{Aho06}:

\noindent
1. $\alpha$ and $\beta$ never derive strings beginning with the same terminal

\noindent
2. At most one of $\alpha$ and $\beta$ can derive the empty string

\noindent
3. If $\beta$ can derive the empty string, then $\alpha$ does not derive any string beginning with a terminal from $FOLLOW(A)$

We use symbol \$ to indicate the end of the input string and $T_{4i}$, $T_{5i}$, $T_{6i}$, $T_{7i}$ as abbreviations for the set of markers, postfixes, prefixes, connectives of priority i, respectively. 

\noindent
\textbf{Proposition 1.} \textit{Tier grammars define LL(1) languages.}

\noindent
\textit{Proof.} $E_i$ cannot derive the empty string. $FIRST(E_i) = \{ T_1, T_2, T_{6i},...,T_{6k} \}$
Clearly, the CFGs in question satisfy the first two LL(1) conditions. To show that the third condition is satisfied as well, we calculate $FIRST$ and $FOLLOW$ sets. We need to consider the following four cases:

1. $G_i \rightarrow \epsilon | s_{i1} | ... | s_{i\overline i} $

\noindent
$FOLLOW(G_i) = \{ T_{71},...,T_{7i-1}, T_{41},...,T_{4q}, T_1, T_2, T_3,$ $T_{61},...,T_{6k}, T_{51},...,T_{5i-1},$ $\$ \}$

2. $ L_i \rightarrow \epsilon | c_{i1} E_{i+1} L_i | ... | c_{i\overline i} E_{i+1} L_i$ (or $L_k \rightarrow \epsilon | c_{k1} C L_k | ... | c_{k\overline k} C L_k$)

\noindent
$FOLLOW(L_i) = \{ T_{71},...,T_{7i-1}, T_{41},...,T_{4q}, T_1, T_2, T_3,$ $T_{61},...,T_{6k}, T_{51},...,T_{5i-1},$ $\$ \}$

3. $ R_i \rightarrow \epsilon | m_{i1} Q_{i+1} R_i | ... | m_{i\overline i} Q_{i+1} R_i$ (or $R_q \rightarrow \epsilon | m_{q1} D R_q | ... | m_{q\overline q} D R_q$)

\noindent
$FOLLOW(R_i) = \{ T_3, T_{41},...,T_{4i-1}, \$ \}$

4. $D \rightarrow \epsilon | E_1 D$

\noindent
$FIRST(E_1 D) = \{ T_1, T_2, T_{61},...,T_{6k} \}$

\noindent
$FOLLOW(D) = \{ T_3, T_{41},...,T_{4q}, \$ \}$

$FIRST$ and $FOLLOW$ do not intersect in all four cases. \qed

The availability of matching LL(1) grammars makes table-driven predictive parsing \cite{Aho06} possible for tier languages. Predictive parsing has a linear time complexity. 

Parse trees for tier grammars are essentially similar to the ASTs of the underlying CFG. One difference is that one node in a tier parse tree combines all associated connectives or markers. Context-free parse trees can be converted to tier parse trees in the following way. Every node $A$ maps to a terminal node. Every node $B$ maps to a Rule 2 node. Every node $E_i$ that corresponds to production $E_i \rightarrow E_{i+1} G_i$ with non-empty $G_i$, or to production $E_i \rightarrow p_{ij} E_i$, or to production $E_i \rightarrow E_{i+1} L_i$ with non-empty $L_i$ maps to Rule 3 node for postfixes, prefixes, or connectives of priority i, respectively. Every node $Q_i$ that corresponds to production $Q_i \rightarrow Q_{i+1} R_i$ with non-empty $R_i$ maps to Rule 4 node for markers of priority i. Nodes $C, F, H, G_i ,L_i, R_i$ are not present in tier parse trees. 

This conversion can be performed by a single traversal of the parse tree built by the CFG predictive parser \cite{Aho06} and better yet can be done during parsing. At the time when a nonterminal is popped from the stack, it is determined whether it remains in the parse tree. If not, this nonterminal is removed from the parse tree and its children are merged with the parent nonterminal.  

Tier languages are unambiguous. Suppose $P_1$ and $P_2$ are any two tier grammar parse trees for the same input string. Since LL(1) languages are unambiguous, $P_1$ and $P_2$ are generated from the same CFG parse tree by applying the same algorithm. Therefore, $P_1$ and $P_2$ are the same. The uniformity of tier languages with respect to predictive parsing is an essential benefit because most questions about properties of CFGs are undecidable. Note that $S \Rightarrow^* N$ for every nonterminal $N$ from tier language parse trees. This is an indication of the inclusiveness of tier grammars.

LL(1) parsing does not require any parser generator tools. A parser can be implemented as a couple of library functions like these Java functions:
\begin{small}
\begin{verbatim}
LL1Parser buildParser(Map<String, List<Set<String>>> terminalGroups);
ParseTree parse(LL1Parser parser, LexemeStream stream);
\end{verbatim}
\end{small}
It is assumed here that terminal classes are mapped to a list of terminal groups, and this list is ordered according to priorities. In the case of gigantic documents, parsing can be implemented via callbacks like it is done in the SAX API for XML in Java (\begin{small}\texttt{http://www.saxproject.org}\end{small}):
\begin{small}
\begin{verbatim}
void parse(LexemeStream stream, EventHandler handler);
\end{verbatim}
\end{small}
where class $EventHandler$ has callback methods for terminals from $T_1, T_2, T_3$, as well as for prefixes, postfixes, connectives, markers. The latter methods are called when the corresponding nonterminal is popped from the stack.

\section{Examples and Applications}

Typical data dump formats such as CSV and other formats for multidimensional arrays can be easily specified as tier grammars. The same applies to the output of many Unix commands and of many command-line tools. Two examples of data formats that can be parsed with using tier grammars are presented in \cite{Sakharov15}: BibTex, documents with numbered sections in which empty lines separate paragraphs.

Machine-generated human-readable files are the main source of examples of tier languages. The output of Apache's \begin{small}\texttt{ReflectionToStringBuilder}\end{small} is one example (\begin{small}\texttt{http://commons.apache.org}\end{small}). Stack traces give other examples. Let us look at some code fragments that generate log files. These code patterns demonstrate why log files or their parts are usually tier languages. The following pseudo-code is self-explanatory. 

\begin{small}
\begin{verbatim}
print(<opening bracket>); loop: { ... print(<data>); ... } 
   print(<closing bracket>);
function f(...){ print(<opening bracket>); ... f(...); ... 
   print(<closing bracket>); return; }	
loop: { ... case ...: print(<prefix>); print(<data>); ... }	
loop: { ... print(<data>); if ( ... ) print(<postfix>); ... }	 
loop: { ... print(<data>); print(<connective>); print(<data>); ... }
loop: { if ( !first ) print(<connective>); ... print(<data>); ... }
loop: { loop: { loop: { ... print(<data>); 
   ... } ... print(<high priority marker>); 
   ... } ... print(<low priority marker>); ... }
\end{verbatim}
\end{small}

The availability of ASTs for data opens up multiple opportunities for other applications. These other applications include:

\begin{itemize}
\item Transformation into XML or JSON

\item Loading data into relational or NoSQL databases

\item Querying documents combining NL fragments, numbers, and codes

\item Publishing web sites (multiple inter-linked HTML pages)

\item Knowledge extraction (in the form of RDF, for example)

\item Preprocessing (data munging) for business intelligence 
\end{itemize}

\section{\uppercase{Analysis}}

Normally, the role of new grammar notations is to introduce richer sets of formal languages or sets with better algebraic properties. By contrast, the role of tier grammars is to provide rich and meaningful parse trees for data. The set of tier languages is a proper subset of LL(1) languages. It includes languages that are not regular. For instance, the Dyck languages over alphabets with one element are tier languages, but they are not regular languages \cite{Berstel02}. Since tier languages are designed to be as inclusive as possible, they do not even include some restrictive regular languages. For instance, the language defined by regular expression $(ab)^*$ and any language with a finite set of distinct strings are not tier languages.  

Yet another aspect of tier grammars is that they can be represented as regular CFGs with one nonterminal. Regular CFG are defined in \cite{Berstel02} and should not be confused with regular grammars. The right-hand sides of productions of regular CFGs include regular expressions over terminals and nonterminals. This alternative representation does not give proper parse trees, however. The right-hand-side (R) of the single production of a matching regular CFG is built by the following recursive process.

\noindent
Step 1. $R := ( b_1 | ... | b_{\underline b} | ( r_1 | ... | r_{\underline r} ) S ( e_1 | ... | e_{\underline e} ) )$

\noindent
Step 2. For $i=k,...,1$, do: 

$R := ( R ( \epsilon | s_{i1} | ... | s_{i\underline i} ) )$ (postfix)

$R := ( ( p_{i1} | ... | p_{i\underline i} )^* R )$ (prefix)

$R := ( R ( ( c_{i1} | ... | c_{i\underline i} ) R )^* )$ (connective)

\noindent
Step 3. $R := R^*$

\noindent
Step 4. For $i=q,...,1$, do:
$R := ( R ( ( m_{i1} | ... | m_{i\underline i} ) R )^* )$

The proof that the matching regular CFG defines the same language is straightforward and done by induction on the total number of marker, prefix, postfix, and connective priorities. In the induction step, the lowest priority is removed in order to use the induction assumption, and then, it is added back. This representation shows that if a tier grammar does not have brackets, then it defines a regular language. In the absence of brackets, R does not contain nonterminals. 

Since the tier grammar notation does not involve any kind of formulas, terminals can only serve as tags giving a particular syntactic meaning to neighboring items or to strings starting or ending with them. Prefixes give a syntactic meaning to the item to the right. Postfixes do the same for the item to the left. A connective glues together the two items adjacent to it. Markers group items on the left and on the right. Brackets define construct borders. Given the design constraints of tier grammars, markers, connectives, prefixes, postfixes, and brackets cover more important cases.

Now we present a simple characterization of tier language strings to show that every tier language includes a wide variety of strings. This guarantees that most data formats can be parsed. The following proposition gives simple conditions for determining whether a string belongs to a tier language. All conditions except one are local - they are checked for pairs of consecutive tokens. One corollary of this proposition is that all strings belong to every tier language containing only base terminals and markers.

\noindent
\textbf{Proposition 2.} \textit{A string belongs to a given tier language if and only if the following conditions hold:}

\noindent
\textit{- brackets are balanced, i.e. the number of opening brackets in the string is equal to the number of closing brackets, and the number of opening brackets is greater than or equal to the number of closing brackets in any prefix substring}

\noindent
\textit{- every postfix follows a base token, closing bracket, or another postfix of a higher priority}

\noindent
\textit{- every prefix precedes a base token, opening bracket, or prefix of the same or higher priority}

\noindent
\textit{- every connective follows a base token, closing bracket, or postfix of a higher priority and precedes a base token, opening parenthesis, or prefix of a higher priority}

\noindent
\textit{Proof.} Let us prove the 'only if' part first. The proof of the first condition is done by induction on the number of parse tree nodes labeled by production $B \rightarrow F S H$ in the derivation of a given string. 

Base. No nodes are labeled by this production. Clearly, the first condition is satisfied.

Induction step. Consider an innermost node labeled by this production. Since the parent node of $B$ is always $C$, we can replace this node and all its descendants by $A \rightarrow t_{11}$. The modified tree also represents a valid derivation. By the induction assumption, the first condition holds for the input string of the modified derivation. Note that the derivation of $S$ from the right-hand side of the production under consideration does not contain any terminals from $T_2$ or $T_3$. Therefore, the first condition holds for the original string.

If $LAST$ is defined similar to $FIRST$, then 
$FIRST(E_i) = \{ T_1, T_2, T_{6i},...,$ $T_{6k}\}$,
$LAST(E_i) = \{ T_1, T_3, T_{5i},...,T_{5k} \}$.
Postfixes of priority $i$ follow $E_{i+1}$ in all derivations. No $E_i$ can derive the empty string. Prefixes of priority $i$ always precede $E_i$. Connectives of priority $i$ always follow and preceed $E_{i+1}$. These observations prove the remaining three conditions.

The 'if' part is proved by triple induction on the number of bracket pairs, on the number of markers, and on the total number of postfixes, prefixes, and connectives. In each induction step, we remove one postfix, prefix, connective, marker, or one pair of brackets from the input string, and show that the conditions remain intact after the removal. The modified string belongs to the tier language in question by the induction assumption. After that, we alter the derivation for the modified string so that it results in the original string. 

Base 1: no brackets. It is proved by induction on the number of markers.

Base 2: no markers. It is proved by induction on total number of postfixes, prefixes, connectives.

Base 3: no postfixes, prefixes, connectives. In this case, the input is a sequence of base tokens which is a valid tier language string.

Induction step 3 (postfix/prefix/connective). Consider terminal $t$ that is a postfix, prefix, or connective of the highest priority $k$.

1. $t$ is a postfix. It follows base token $b$. Let us remove $t$ from the input string. The conditions remain intact after the removal. The modified string belongs to the tier language in question by the induction assumption. The derivation of the modified string must be as follows:

\noindent
$S \Rightarrow^* ... E_k ... \Rightarrow ... C G_k ... \Rightarrow ... C ... \Rightarrow^* ... b ...$

Replace it with another valid derivation of the original string:

\noindent
$S \Rightarrow^* ... E_k ... \Rightarrow ... C G_k ... \Rightarrow ... C t ... \Rightarrow^* ... b t ...$ 

2. $t$ is a prefix. It precedes base token $b$ or another prefix $p$. Consider the innermost $t$. Let us remove $t$ from the input string. The conditions remain intact after the removal. The derivation of the modified string must be:

\noindent
$S \Rightarrow^* ... E_k ... \Rightarrow ... C ... \Rightarrow^* ... b ...$

Replace it with another valid derivation of the original string:

\noindent
$S \Rightarrow^* ... E_k ... \Rightarrow^* ... t C ... \Rightarrow^* ... t b ...$

3. $t$ is a connective. It follows base token $b_1$ and precedes base token $b_2$. Consider the first occurrence of $t$. Let us remove it along with $b_2$. The conditions remain intact. The derivation of the modified string must be one of these two:

\noindent
$S \Rightarrow^* ... E_k ... \Rightarrow ... C L_k ... \Rightarrow ... C ... \Rightarrow^* ... b_1 ...$

\noindent
$S \Rightarrow^* ... E_k ... \Rightarrow ... C L_k ... \Rightarrow^* ... b_1 L_k ...$

In the first case, replace it with another valid derivation of the original string:

\noindent
$S \Rightarrow^* ... E_k ... \Rightarrow ... C L_k ... \Rightarrow ... C t C L_k ... \Rightarrow^* ... b_1 t b_2 ...$

In the second case, replace it with another valid derivation of the original string:

\noindent
$S \Rightarrow^* ... E_k ... \Rightarrow ... C L_k ... \Rightarrow ... C t C L_k ... \Rightarrow^* ... b_1 t b_2 L_k ...$

Induction step 2 (markers). Let us remove the leftmost marker $m$ of the highest priority q. The conditions remain intact. The derivation of the modified string is either $S \Rightarrow^* ... Q_q ... \Rightarrow^* ... D ... \Rightarrow^* ... E_1 ... E_1 ...$ or $S \Rightarrow^* ... Q_q ... \Rightarrow ... D R_q ... \Rightarrow^* E_1 ... E_1 R_q$.
The proposition conditions imply that a marker cannot follow a prefix or connective, and cannot precede a postfix or connective. Note that in absence of brackets, no $E_1$ derivation contains two adjacent base tokens, prefixes can only be preceded by connectives, and postfixes can only be followed by connectives within any $E_1$ derivation. Therefore the position of the removed marker m can only be between two consecutive $E_1$, or before the first $E_1$, or after the last $E_1$. We can replace the first derivation of the modified string with a valid derivation of the original string:

\noindent
$S \Rightarrow^* ... Q_q ... \Rightarrow ... D R_q ... \Rightarrow^* ... E_1 ... E_1 m E_1 ... E_1$

And we can replace the second derivation with another valid derivation of the original string:

\noindent
$S \Rightarrow^* ... Q_q ... \Rightarrow ... D R_q ... \Rightarrow^* ... E_1 ... E_1 m E_1 ... E_1 R_q ...$

Induction step 1 (brackets). Let us replace the first innermost pair of brackets $p_1$, $p_2$ along with everything in between with base token $b$. The conditions remain intact. Consider the derivation of the modified string: 

\noindent
$S \Rightarrow^* ... C ... \Rightarrow ... A ... \Rightarrow ... b ...$

String $u$ comprised of the sequence of tokens within the removed brackets satisfies the proposition conditions. By the proof of the base case, $S \Rightarrow^* u$. We can replace the above derivation with another valid derivation of the original string:

\noindent
$S \Rightarrow^* ... C ... \Rightarrow^* ... B ... \Rightarrow^* ... p_1 S p_2 ...$
\qed

\textbf{Corollary.} \textit{If $T'_2$, $T'_3$, $T'_4$, $T'_5$, $T'_6$, $T'_7$ are subsets of $T_2$, $T_3$, $T_4$, $T_5$, $T_6$, $T_7$, respectively, $s \in \Lambda(\{T_1$, $T_2$, $T_3$, $T_4$, $T_5$, $T_6$, $T_7\})$, and terminals from $T_2 \setminus T'_2$ are balanced with terminals from $T_3 \setminus T'_3$ in $s$, then $s \in \Lambda(\{T_1 \cup T'_2 \cup T'_3 \cup T'_4 \cup T'_5 \cup T'_6 \cup T'_7, T_2 \setminus T'_2, T_3 \setminus T'_3, T_4 \setminus T'_4, T_5 \setminus T'_5, T_6 \setminus T'_6, T_7 \setminus T'_7\})$.}

This corollary guarantees that parsing with incomplete syntax will work. The extension of syntax usually amounts to assigning other roles to some of the base terminals.

\section{Related Work}

Several alternatives to the notation of CFGs have been developed. All these alternatives are similar to CFGs in terms of the complexity of creating and debugging grammars. One such alternative notation is parsing expression grammars \cite{Ford04}. The difference between parsing expression grammars and traditional CFGs is in the interpretation of grammar rules.  Operator precedence grammars can be considered an alternative notation too. They are defined by the operator precedence matrix \cite{Meduna07}, which is tricky to create. Even some heuristic rules were suggested to assist people with this task, as there is no explicit connection between the matrix and parse trees. Balanced grammars \cite{Berstel02} combine productions with regular expressions over terminals and nonterminals. Their parse trees are sparse because of the use of regular expressions in grammar productions. 

Visibly pushdown grammars \cite{Alur04} are one example of notations that are derivative from that of CFGs. Attribute grammars \cite{Aho06} are an example of CFG extensions. Stochastic CFG parsers \cite{Chappelier98} have a prohibitive time complexity for data that may be much bigger than programs. The complexity of CFGs has driven interest in the approximation of context-free languages with regular languages \cite{Egecioglu09}. 

Despite the remarkable research in the area of formal grammars, its applications to data parsing are few and far between. The use of attribute grammars for specification and parsing binary file formats \cite{Underwood12} is one example. PacketTypes \cite{McCann00} and DataScript \cite{Back02} are examples of data description languages that introduce complex notations that are quite different from CFGs. These languages handle binary data. Tier grammars are designed for character data (ASCII or Unicode). An overview of data description languages can be found in \cite{Fisher06}. 

PADS \cite{Fisher05} is a particularly sophisticated and more general data description language. It even supports automatic grammar inference \cite{Fisher08}. ANNE \cite{Xi10} is an eclectic tool that derives PADS \cite{Fisher05} data format specifications from user-annotated data sources. Data Format Description Language (DFDL) \cite{Powell11} is a complex language for the specification of data formats. DFDL is based on the XML Schema. None of these data description languages are on par with tier grammars in terms of simplicity of specification of data formats. 

Grammar inference methods are basically limited to regular languages and other simple languages \cite{Sakakibara97}. RoadRunner \cite{Crescenzi04} infers union-free regular grammars that are used to extract information from large web sites. A method of learning CFG productions that specify the syntax of web server access logs is presented in \cite{Thakur13}. The log format considered in this paper is a very simple regular language. It is not clear if this inference method will work for more complex languages. 

\section{Conclusion}

Grammars enable the declarative programming of data parsers. Specifying a grammar by splitting terminals into meaningful disjoint subsets is one of the easiest ways to describe syntax. It is even simpler than regular expressions. The family of tier grammars presented and investigated here has sufficient expressive power to describe the syntax of many data languages. Tier grammars have the qualities that are important for data parsing, particularly for parsing big data. The idea behind tier grammars that leads to LL(1) conditions is considering nonterminals as a set ordered by respective priorities, and limiting productions to the forms in which forward references in the right-hand sides are always to the next nonterminal and backward references are bracketed by terminals. 

If the expressiveness of tier grammars is not sufficient, they can be easily extended. One extension is the addition of prefixes of arity more than one. The relevant context-free productions are $E_i \rightarrow p_i E_i ... E_i$ where the number of $E_i$ in the right-hand side is the arity of $p_i$. Also, multiple tier grammars can be combined so that every source grammar applies only to a relevant portion of a document. The advantage of combining multiple tier grammars vs CFGs is that the simplicity of the notation is not compromised. These extended and combined grammars remain LL(1). For more information about extending and combining tier grammars, see \cite{Sakharov15}. Tier grammars can be further generalized, and a description of such generalization will be published separately.

\bibliographystyle{splncs}
{\small
\bibliography{TierGrammarsArxiv}}

\begin{thebibliography}{10}

\bibitem{Aho06}
Aho, A.V., Lam, M.S., Sethi, R., Ullman, J.D.:
\newblock Compilers: Principles, Techniques, and Tools (2nd Edition).
\newblock Addison-Wesley Longman Publishing Co., Inc., Boston, MA, USA (2006)

\bibitem{Sakharov15}
Sakharov, A., Sakharov, T.:
\newblock Data parsing using tier grammars.
\newblock In: Proceedings of the 7th International Joint Conference on
  Knowledge Discovery, Knowledge Engineering and Knowledge Management, Lisbon,
  Portugal. (2015)  463--468

\bibitem{Berstel02}
Berstel, J., Boasson, L.:
\newblock Balanced grammars and their languages.
\newblock In: Formal and Natural Computing - Essays Dedicated to Grzegorz
  Rozenberg [on occasion of his 60th birthday, March 14, 2002]. (2002)  3--25

\bibitem{Ford04}
Ford, B.:
\newblock Parsing expression grammars: A recognition-based syntactic
  foundation.
\newblock In: Proceedings of the 31st ACM SIGPLAN-SIGACT Symposium on
  Principles of Programming Languages. POPL '04, New York, NY, USA, ACM (2004)
  111--122

\bibitem{Meduna07}
Meduna, A.:
\newblock Elements of Compiler Design. 1st edn.
\newblock Auerbach Publications, Boston, MA, USA (2007)

\bibitem{Alur04}
Alur, R., Madhusudan, P.:
\newblock Visibly pushdown languages.
\newblock In: Proceedings of the Thirty-sixth Annual ACM Symposium on Theory of
  Computing. STOC '04, New York, NY, USA, ACM (2004)  202--211

\bibitem{Chappelier98}
Chappelier, J.C., Rajman, M.:
\newblock A generalized cyk algorithm for parsing stochastic cfg.
\newblock In: Proceedings of Tabulation in Parsing and Deduction (TAPD'98),
  Paris, France (1998)  133--137

\bibitem{Egecioglu09}
Egecioglu, {\"{O}}.:
\newblock Strongly regular grammars and regular approximation of context-free
  languages.
\newblock In: Developments in Language Theory, 13th International Conference,
  {DLT} 2009, Stuttgart, Germany, June 30 - July 3, 2009. Proceedings. (2009)
  207--220

\bibitem{Underwood12}
Underwood, W.:
\newblock Grammar-based specification and parsing of binary file formats.
\newblock International Journal of Digital Curation \textbf{7}(1) (2012)
  95--106

\bibitem{McCann00}
McCann, P.J., Chandra, S.:
\newblock Packet types: Abstract specification of network protocol messages.
\newblock In: Proceedings of the Conference on Applications, Technologies,
  Architectures, and Protocols for Computer Communication. SIGCOMM '00, New
  York, NY, USA, ACM (2000)  321--333

\bibitem{Back02}
Back, G.:
\newblock Datascript - a specification and scripting language for binary data.
\newblock In: In Generative Programming and Component Engineering, Springer
  (2002)  66--77

\bibitem{Fisher06}
Fisher, K., Mandelbaum, Y., Walker, D.:
\newblock The next 700 data description languages.
\newblock In: Conference Record of the 33rd ACM SIGPLAN-SIGACT Symposium on
  Principles of Programming Languages. POPL '06, New York, NY, USA, ACM (2006)
  2--15

\bibitem{Fisher05}
Fisher, K., Gruber, R.:
\newblock Pads: A domain-specific language for processing ad hoc data.
\newblock In: Proceedings of the 2005 ACM SIGPLAN Conference on Programming
  Language Design and Implementation. PLDI '05, New York, NY, USA, ACM (2005)
  295--304

\bibitem{Fisher08}
Fisher, K., Walker, D., Zhu, K.Q.:
\newblock Learnpads: Automatic tool generation from ad hoc data.
\newblock In: Proceedings of the 2008 ACM SIGMOD International Conference on
  Management of Data. SIGMOD '08, New York, NY, USA, ACM (2008)  1299--1302

\bibitem{Xi10}
Xi, Q., Walker, D.:
\newblock A context-free markup language for semi-structured text.
\newblock In: Proceedings of the 31st ACM SIGPLAN Conference on Programming
  Language Design and Implementation. PLDI '10, New York, NY, USA, ACM (2010)
  221--232

\bibitem{Powell11}
Powell, A., Beckerle, M., Hanson, S.:
\newblock Data format description language (dfdl).
\newblock Technical report, Open Grid Forum (1 2011)

\bibitem{Sakakibara97}
Sakakibara, Y.:
\newblock Recent advances of grammatical inference.
\newblock Theoretical Computer Science' \textbf{185}(1) (October 1997)  15--45

\bibitem{Crescenzi04}
Crescenzi, V., Mecca, G.:
\newblock Automatic information extraction from large websites.
\newblock J. ACM \textbf{51}(5) (September 2004)  731--779

\bibitem{Thakur13}
Thakur, R., Jain, S., Chaudhari, N.S.:
\newblock User behavior analysis using alignment based grammatical inference
  from web server access log.
\newblock International Journal of Future Computer and Communication
  \textbf{2}(6) (2013)  543

\end{thebibliography}

\end{document}